\documentclass[%
 reprint,
 amsmath,amssymb,
 aps,
prb,
]{revtex4-1}

\usepackage{graphicx}
\usepackage{dcolumn}
\usepackage{bm}
\usepackage{braket}
\usepackage{caption}
\usepackage{subcaption}
\usepackage{hyperref}

\usepackage[usenames,dvipsnames]{color}

\newcommand{\bi}{\bibitem}
\newcommand{\cc}{\captionsetup{justification=raggedright,singlelinecheck=false}}

\newcommand{\ct}{\cite}

\newcommand{\vk}{\vec{k}}
\newcommand{\ek}{\epsilon(k)}

\begin{document}

\title{Role of topology on the work distribution function of a \\quenched Haldane  model of graphene}

\author{Sourav Bhattacharjee, Utso Bhattacharya and Amit Dutta\\
Department of Physics, Indian Institute of Technology Kanpur-208016, India}

\date{\today}

\begin{abstract}
We investigate the effect of equilibrium topology on the statistics of non-equilibrium work performed during the subsequent unitary evolution,  following a sudden quench of the Semenoff mass of the Haldane model. We show that  the resulting work distribution function for quenches performed on the Haldane Hamiltonian with broken time reversal symmetry (TRS)  exhibits richer universal characteristics as compared to  those performed  on the time-reversal symmetric massive graphene limit whose work distribution function we have also evaluated for comparison. Importantly, our results show that the work distribution function exhibits different universal behaviors following the non-equilibrium dynamics of the system for small $\phi$ (argument of complex next
nearest neighbor hopping) and large $\phi$ limits, although the two limits belong to the same equilibrium universality class.
\end{abstract}

\maketitle

\section{\label{sec:level1}Introduction}
Studying the probability distribution of work in a driven quantum system is an interesting area of recent research \ct{silva08, paran09, gambassi_ar_11, gambassi11, gambassi12, sotiriadis13, smacchia13,  dorner13,  mazzola13, batalhao14, sindona14, shchadilova14, palmai14, fusco14, palmai15, solinas15, sengupta15,  bayocboc15, russomanno15, santos16, talarico16, lobejko17, rotondo18, goold18}.
We recall that in quantum mechanics, work ($W$) is not an observable, rather it  acquires a stochastic behavior due to the inherent probabilistic nature of quantum measurements \ct{campisi11, talkner07}. Naturally, the object of interest therefore is no longer $W$ itself but rather a distribution function $P(W)$ which encodes its fluctuating behaviour. This work distribution function is also intricately  connected with popular information theoretic tools like fidelity, fidelity susceptibility and Loschmidt echo \cite{quan06, zhou08, venuti11}. The motivation behind studying the work distribution function, particularly for many body systems, lies in understanding the non-linear responses embedded in the fluctuation relations given in terms of work, heat and entropy. Such understanding is crucial in light of the emerging field of quantum thermodynamics where recent research has focused on identifying the principles from which the known thermodynamic laws in the macroscopic limit can be derived  \cite{scovil59, scully03, allah04, erez08, horodecki13, skrz14, brandao15, uzdin15}. Additionally, the progressing miniaturization of physical devices to scales where quantum effects become dominant has raised the question whether the well known efficiency limits for work extraction in macroscopic thermodynamics holds in the quantum regime \cite{alicki79, correa14, pekola15, campisi16, robnagel16, wolfgang18} (see \cite{kosloff13, gelb15, goold16, vinj16, kosloff17, alicki18} for review). On the other hand, numerous works on the topological aspects of statistical mechanics have led to a growing evidence that topology has a profound effect on both the equilibrium \cite{haldane88, kane05} and non-equilibrium \cite{patel13, thakurathi13, wang17, tarnowski, caio15, caio16} dynamics of a given system. As such, it is imperative to explore any possible effect that a system's topological structure might have on the work distribution function.

Remarkably, it has been shown that the work distribution function attains a universal behavior \ct{gambassi_ar_11, gambassi11, gambassi12, sotiriadis13, smacchia13} following a quench of the Hamiltonian of the system in the vicinity of a quantum critical point (QCP). Moreover, $P(W)$ displays an interesting  edge behaviour \ct{ gambassi_ar_11,  sotiriadis13,  smacchia13} following a gap in the small $W$($W \to 0$) limit with a power-law behavior with $W$ and  the associated exponent depends on the initial and final value of of the quench parameter (with respect to the critical point) and the spatial dimensionality. This universal behavior  has been probed  extensively in free bosonic as well as free fermionic models for
both global and local quenches. 

Generally, to define the amount of work done on the system as a
result of the sudden quench performed, one must make two projective
measurements. Considering the system to be initially in thermal equilibrium, the first measurement projects onto the eigenbasis of
the initial Hamiltonian $H_i$ at $t = 0$
with probability $p^0_n = e^{- \beta E_n^0}/Z_i$ (where $\beta$ is the inverse temperature, $E_n^0$ are the energy eigenvalues and $Z_i$ is the initial partition function of the system). Following a sudden quench, the system evolves freely till a time $\tau$ after which the second projective measurement is carried out onto the eigenbasis $\ket{\phi^n_\tau}$ of the final Hamiltonian,
with a probability $p_n^\tau = |\langle \phi^n_\tau|\psi_\tau\rangle|^2$. The
fluctuations in the work performed, which is encoded in $P(W)$, therefore arise from both the thermal statistics $p_n^0$ and the quantum measurement statistics 
$p_\tau^n$ over many ensembles. However, in our work, we will ignore the thermal fluctuations and assume that the system is initially prepared in a pure state. Interestingly, the first moment $\langle W\rangle$ of the distribution which is the average work done is exactly equal to the residual energy accumulated during the driven unitary evolution  which in turn serves as a fundamental measure facilitating the understanding of the emergence of steady state behavior in periodically driven many body quantum systems \cite{russo12}.

To elaborate further, let us assume that a closed $d$-dimensional quantum many body system is initially prepared in the ground state $\ket{\psi_0}$ of an initial Hamiltonian $H_i$;   a certain parameter of the Hamiltonian is then quenched at time $t=0$ using some protocol following which the system is allowed to evolve unitarily. 
 The work distribution function $P(W)$ characterising the probability that $W$ amount of work has been done after the the system evolves freely for a time $\tau$  is 
\begin{equation}\label{eq_pw} 
P(W)=\sum_n\delta\left(W-[E^n_\tau-E^0_i]\right)|\braket{\phi^n_\tau|\psi_\tau}|^2
\end{equation}
where $\ket{\psi_\tau}$ is the evolved state of the system at time $\tau$, $\ket{\phi_\tau^n}$ and $E^n_\tau$ denote the $n^{th}$ instantaneous energy eigenstate and  its eigen energy respectively while $E^0_i$ is the (ground state) energy of $\ket{\psi_0}$. If the quench is performed suddenly, the subsequent time evolution of $\ket{\psi_0}$ is  dictated by the final time independent Hamiltonian $H_f$ with the final value of the quench parameter, i.e. $\ket{\psi_t}=e^{-iH_ft}\ket{\psi_0}$. One immediately finds, 
\begin{equation}\label{eq_weight}
|\braket{\phi^n_\tau|\psi_\tau}|^2=|\bra{\phi^n_f}e^{-iH_f\tau}\ket{\psi_0}|^2=|\braket{\phi^n_f|\psi_0}|^2
\end{equation}
where {$ \ket{\phi_f^n}$ are the instantaneous energy eigenstates of $H_f$.
It is now straightforward to show that
\begin{equation}\label{eq_pw_fourier}
P(W)=\int_{-\infty}^{\infty}e^{iW\tau}G(\tau)d\tau
\end{equation}
where $G(\tau)$ is the characteristic function of $P(W)$ and is given as 
\begin{equation}\label{eq_gt}
G(\tau)=e^{-i\Delta E_0\tau}\bra{\psi_0}e^{i(E_f^0-H_f)\tau}\ket{\psi_0}.
\end{equation}
Here $\Delta E_0=E_f^0-E_i^0$ is the difference in ground state energies of the final and initial Hamiltonians and hence is the minimum threshold of possible work. This threshold is set by the adiabatic limit of time evolution implying that the irreversible work $W_{irr} = W - \Delta E_0$ can take only positive values. We also note in passing that upon rescaling $E_f^0$ to zero, the inner product term in Eq.~\eqref{eq_gt} reduces to the conventional Loschmidt overlap amplitude \ct{heyl13}. 

It is interesting to note that an expression similar to that of the characteristic function of the work distribution also arises while calculating the core hole Green's function usually analyzed in the context of X-ray Fermi Edge singularities \cite{mahan67, nozieres69}, which in turn shares a deep connection with the Anderson orthogonality catastrophe problem (AOCP) \cite{anderson67}. Remarkably, Anderson established that the non-interacting ground states become orthogonal as the system size increases with a power-law that depends universally on the phase shift induced by the scattering potential. The calculation of the core hole Green's function involves the determination of the vacuum persistence amplitude (VPA), which is again nothing but the complex conjugate of the characteristic function of work.  Moreover, the absorption spectrum obtained in X-ray scattering experiments, which is the Fourier transform of the VPA, displays a power-law threshold singularity or Fermi Edge singularity due to the power-law decay of VPA. Naturally, one expects such edge singularities to appear in the work distribution function as well which is again the Fourier transform of the characteristic function. Due to the orthogonality catastrophe, the two ground-states before and after the sudden addition (or quench) of the impurity potential, becomes orthogonal in the thermodynamic limit; ensuring that the probability of doing adiabatic work goes to zero as a power-law. This is exactly what we expect thermodynamically both in case of the X-ray Fermi edge singularity behavior and also in the statistics of work distribution after a sudden quench is performed.

Let us now show how the characteristic function is related to the partition function of a higher-dimensional
statistical model. An analytic continuation to imaginary time $\tau=-iS$ enables us to rewrite Eq.~\eqref{eq_gt} in the following way\ct{gambassi_ar_11}
\begin{subequations}
	\begin{equation}\label{eq_gs}
	G(S)=e^{-S\Delta E_0}Z(S)
	\end{equation}
	\begin{equation}\label{eq_zs}
	Z(S)=\bra{\psi_0}(e^{E_f^0-H_f})^S\ket{\psi_0}
	\end{equation}
\end{subequations}  
where $Z(S)$, in accordance with the quantum to classical correspondence principle, can be interpreted as the partition function of a $(d+1)$-dimensional classical system defined on a strip geometry of width $S$ with boundary states $\ket{\psi_0}$. The associated free energy ($F$) can  be decoupled into  three contributions as follows:
\begin{equation}\label{eq_free_energy}
F=-\log{G(S)}=L^d\left(S\times f_b + 2f_s + f_c(S)\right)
\end{equation}
where $f_b=\Delta E_0/L^d$ is the bulk free energy density, $f_s$ is the surface free energy due to the two boundaries of the strip and hence is independent of its thickness $S$ while $f_c(S)$ is the contribution due to the Casimir interaction between the boundaries which decays to zero for large $S$ \ct{gambassi09}. 
\begin{figure*}
	\centering
	\begin{subfigure}{0.4\textwidth}
		\centering
		\includegraphics[width=\columnwidth]{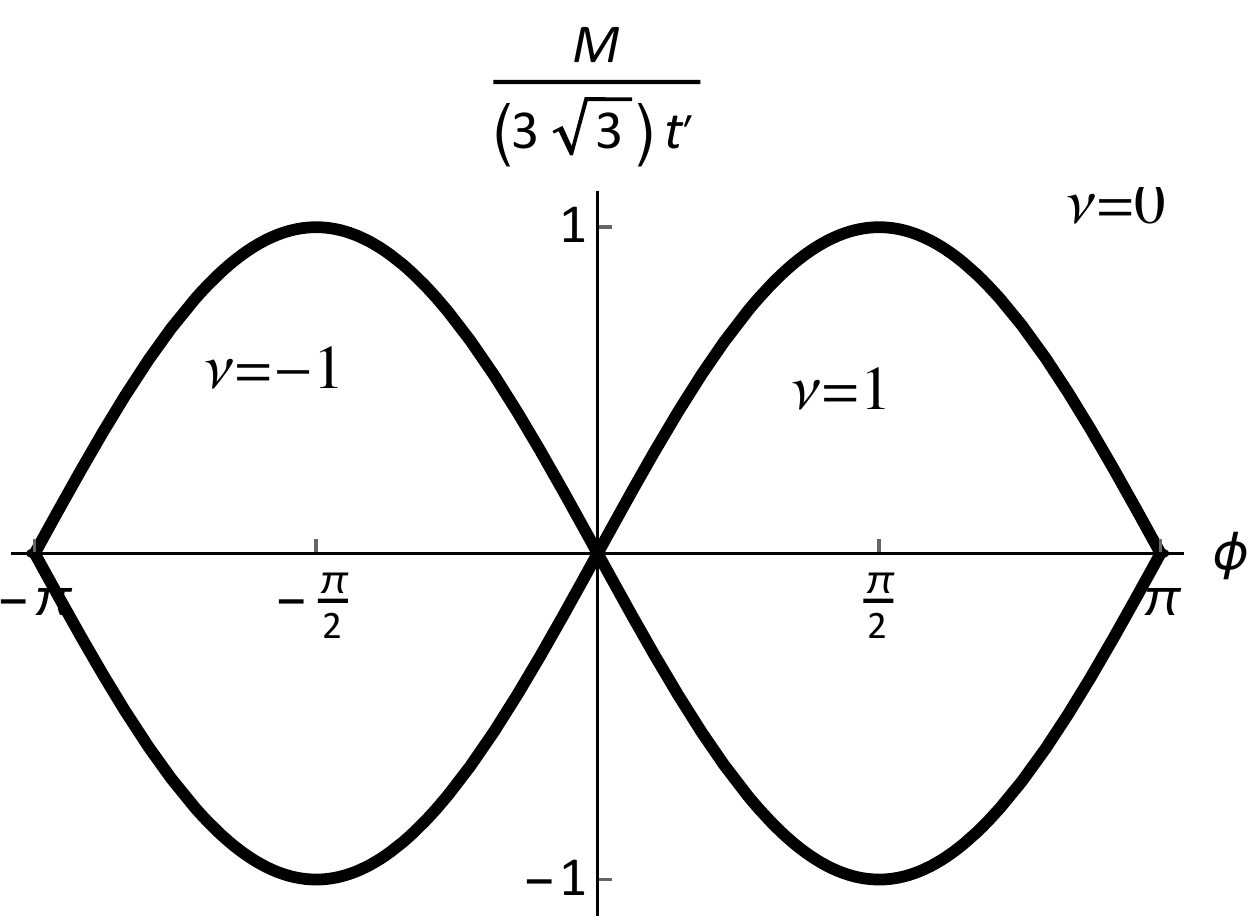}
		\caption{}
		\label{fig_haldane}
	\end{subfigure}\quad\quad\quad\quad
	\begin{subfigure}{0.4\textwidth}
		\centering
		\includegraphics[width=\columnwidth]{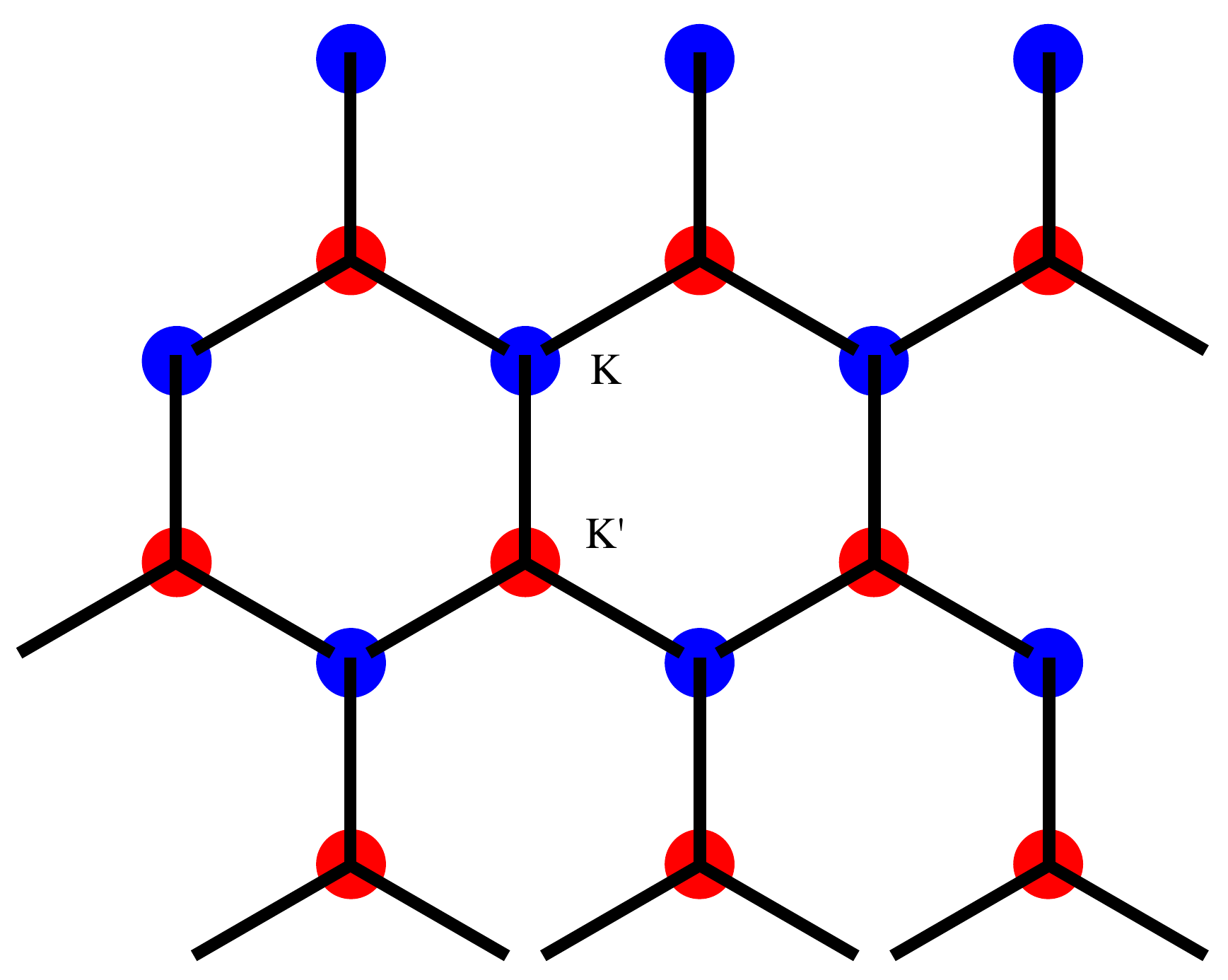}
		\caption{}
		\label{fig_lattice}
	\end{subfigure}
	\cc	
	\caption{(a) Topological phases of Haldane model characterized by integer Chern number $\nu$ values with $t=1$. The two lobes ($\nu=\pm1$) correspond to the phases where robust conducting edge states exist while the area outside the lobes ($\nu=0$) correspond to trivial insulator/conductor phase. The sinusoidal curves are the quantum critical lines (QCLs) which separate the trivial and topological phases. (b) Reciprocal lattice of graphene with only nearest neighbor interactions. The spectrum is gapless at the two sets of inequivalent Dirac points $K$(blue) and $K'$(red) in the absence of any on-site potential. }
\end{figure*}

 Close to a critical point, the response of the system is characterized by a diverging correlation length $\xi$, thus there is a slower non-exponential decay of the two point correlation functions of fluctuations of the order parameter. In such a scenario, the existence of the boundary states impose effective boundary conditions on the order parameter which leads to a Casimir like force between the boundaries. This results in contribution of an additional part $f_c(S)$ (Eq.~\eqref{eq_free_energy}) to the free energy of the system, which assumes the scaling form
\begin{equation}\label{eq_scaling}
f_c(S)=S^{-d} \mathcal{F}(S/\xi);
\end{equation} 
here, $\mathcal{F}(S/\xi)$ is a universal scaling function which is independent of microscopic details and only depends on the surface and bulk universality classes. This is the source of the emergence of universal behaviour of $P(W)$ close to criticality, where the scaling function $\mathcal{F}(S/\xi)$ and hence $f_c(S)$ can be asymptotically expanded for $S/\xi\gg1$. Therefore, the universality in the behavior of $P(W)$ for small $W$ can be extracted from the large $S$ behavior $f_c(S)$. For the rest of the paper, we will only focus on this low work regime of $P(W)$.

Let us now briefly recapitulate some of the generic aspects of the universal behavior of $P(W)$ valid for a wide class of free fermionic models. Especially, focusing on the  $1$-D transverse field Ising model  with the transverse field close to its critical value $g_c$, $P(W)$ depends solely on the relative value of the initial field $g_i$ and the final field $g_f$ (after a sudden quench) with respect to $g_c$ \ct{gambassi_ar_11, smacchia13}. In other words, it depends on whether the quench is carried out within the same quantum phase ($g_i,g_f\gtrless g_c$), or across the quantum phases ($g_i>g_c, g_f<g_c$ or $g_i<g_c, g_f>g_c$), or from (to) the critical point ($g_{i(f)}=g_c$). However, a few characteristics are common in all the cases; there is a delta function peak at the origin with a weight factor given by the ground state fidelity $|\braket{\phi^0_f|\psi_0}|^2$. This corresponds to the reversible work which is the difference of the initial and final ground state energies as discussed above. In addition, there also exists an edge  at a lower cutoff of $W$ below which $P(W)$ is zero.

In this paper, we explore the effect of equilibrium topology on the non-equilibrium work statistics following a sudden quench of a parameter of the system Hamiltonian. This is relevant in the light of a growing number of recent studies which explore connections between equilibrium topology and dynamics, both in the context of periodic \cite{oka09,kitagawa10,lindner11} and quench \cite{patel13,rajak14,caio15,caio16,budich16,sharma16,heyl18,utso17,sougata18} dynamics. We study the non-equilibrium dynamics of the paradigmatic Haldane model \ct{haldane88} which is an integrable two dimensional model of spin-less electrons; the phase diagram of the model (Fig.~\ref{fig_haldane}) hosts topological as well as trivial phases. This model is based on an infinite graphene like honeycomb lattice (Fig.~\ref{fig_lattice}) with broken sub-lattice symmetry (SLS) and time-reversal symmetry (TRS) manifested in nearest neighbor (NN) and complex next-nearest neighbor(NNN) hoppings. The Hamiltonian of the Haldane model can be decomposed as a sum of Hamiltonians of decoupled two-level systems, 
\begin{equation}
H=\sum_{\vec{k}} H(\vec{k})=\sum_{\vec{k}}\vec{h}(\vec{k})\cdot\vec{\sigma}+h_0(\vk)\mathbb{\textit{I}}
\label{eq_hk}
\end{equation}
where $\vec{\sigma}\equiv\left(\sigma_x,\sigma_y,\sigma_z\right)$ are the Pauli matrices, $\mathbb{\textit{I}}$ is the ($2\times2$) identity matrix and
\begin{subequations}
	\begin{equation}
	h_x(\vec{k})=-t\left(\cos{(\vec{k}\cdot\vec{e_1})}+\cos{(\vec{k}\cdot\vec{e_2})}+\cos{(\vec{k}\cdot\vec{e_3})}\right)
	\label{eq_hx}
	\end{equation}
	\begin{equation}
	h_y(\vec{k})=-t\left(\sin{(\vec{k}\cdot\vec{e_1})}+\sin{(\vec{k}\cdot\vec{e_2})}+\sin{(\vec{k}\cdot\vec{e_3})}\right)
	\label{eq_hy}
	\end{equation}
	\begin{equation}
	h_z(\vec{k})=M-2t'\sin{\phi}\left(\sin{(\vec{k}\cdot\vec{v_1})}+\sin{(\vec{k}\cdot\vec{v_2})}+\sin{(\vec{k}\cdot\vec{v_3})}\right)
	\label{eq_hz}
	\end{equation}
	\begin{equation}
	h_0(\vk)=-2t'\cos{\phi}\left(\cos{(\vec{k}\cdot\vec{v_1})}+\cos{(\vec{k}\cdot\vec{v_2})}+\cos{(\vec{k}\cdot\vec{v_3})}\right).
	\end{equation}
\end{subequations}
Here, for a given lattice site, the vectors $\{\vec{e_i}\}$ and $\{\vec{v_i}\}$ ($i=1,2,3$) are the locations of NN and NNN sites respectively. Further, $t$ is the amplitude of NN hopping in the graphene honeycomb lattice,  $t'$ is the absolute part of the complex NNN hopping and $\phi$ is its argument; $M$, on the other hand, denotes the staggered on-site potential at the lattice sites, also known as the Semenoff mass. When $M=t'=0, t=1$, the Hamiltonian reduces to that of the gapless graphene Hamiltonian with no topological properties. 

The topological nature of the Haldane model is an artefact of the simultaneous presence of the Semenoff mass and the complex NNN hoppings in the Hamiltonian which are responsible for breaking the SLS and TRS of the original graphene lattice, respectively \ct{haldane88}. The different topological phases are characterized by a topological order parameter called the Chern number ($\nu$). When $\nu=0$, the system  behaves as a trivial insulator/conductor while for $\nu=\pm1$, conducting edge states arise which are topologically protected and hence robust while the bulk of the system remains insulating; the topological and trivial phases are separated by the quantum critical lines (QCLs).  The phase diagram  is shown in Fig.~\ref{fig_haldane}.

The motivation of our work is therefore to analyze the effect of the above mentioned topological structure on the work statistics of the system. To achieve this goal, we first perform quenches on the Semenoff mass $M$ fixing $\phi=0$ (so that TRS is intact) and elucidate the dependence of $P(W)$ on the initial value $M_i$ and final value $M_f$ of the Semenoff mass. In this case, the quench is always performed in the topologically trivial state for any $M_i$ and $M_f$. We then proceed to the case with small $\phi\neq 0$ (so that the model now has a non-trivial topology Fig.~\ref{fig_haldane}) and perform similar quenches in the vicinity of QCLs and highlight the interesting features appearing in $P(W)$.

Our results are summarized at the outset as follows. We find that the universal nature of P(W) in the case of quenches in $M$ performed on the massive graphene Hamiltonian depends on the relative position of $M_i$ and $M_f$ with respect to the critical gapless point $M=0$. However, when the quenches are performed in the TRS broken topological Haldane Hamiltonian, we observe completely new and rich behavior. This new behavior thus emerges only when both SLS and TRS are broken and therefore is a consequence of the resulting topological structure of the model.

The rest of the paper is organised as follows. In Sec.~\ref{sec_casimir}, we review the procedure for calculating the critical Casimir free energy. In Sec.~\ref{sec_triv} and Sec.~\ref{sec_topo}, the work distribution function is calculated for quenches in the $\phi =0$ and  small $\phi\neq 0$, respectively. The discussions and concluding comments are presented in Sec.~\ref{sec_dis} and the experimental possibilities are discussed in Sec.~\ref{sec_exp}. We further present two appendices showing small momentum expansion (Appendix \ref{app_1}) and the Mellin transform (Appendix \ref{app_2}) approach for evaluating $P(W)$.
\begin{figure*}
	\begin{subfigure}{0.2\textwidth}
		\includegraphics[width=0.8\columnwidth]{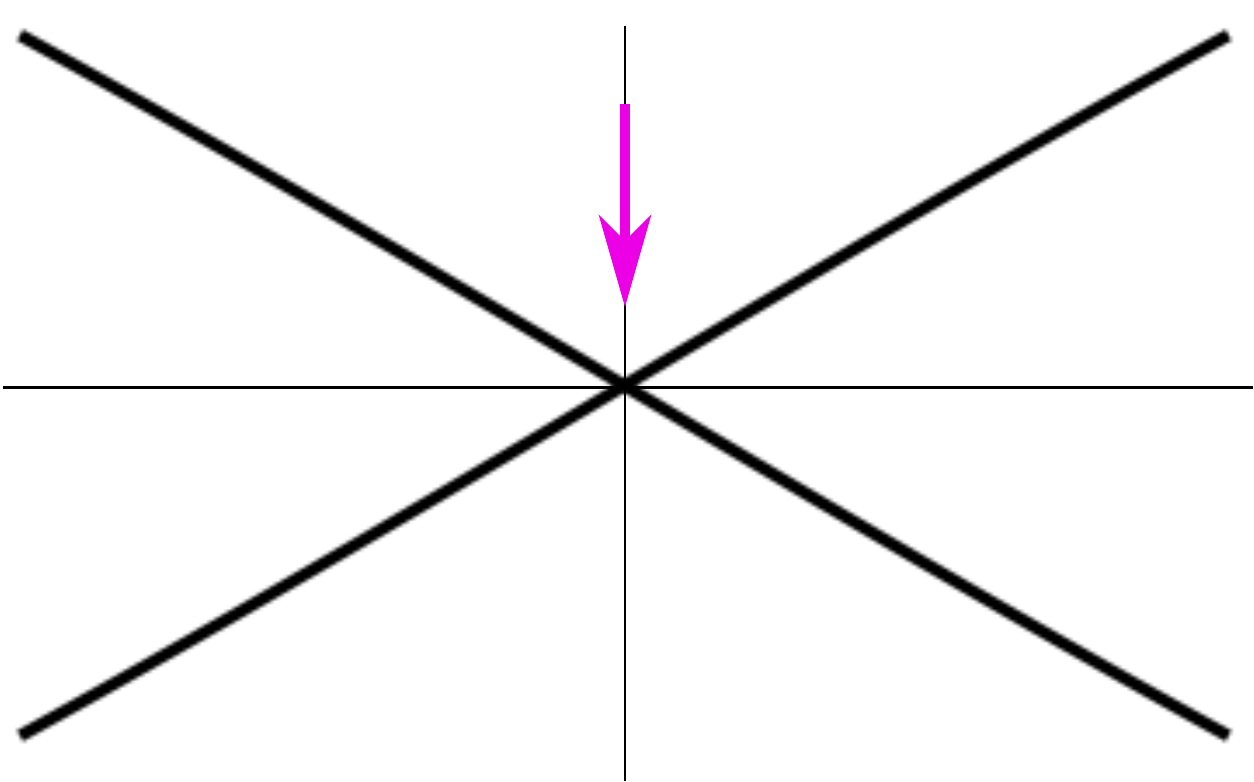}
		\caption{}
		\label{fig_quench_1}
	\end{subfigure}
	\begin{subfigure}{0.2\textwidth}
		\includegraphics[width=0.8\columnwidth]{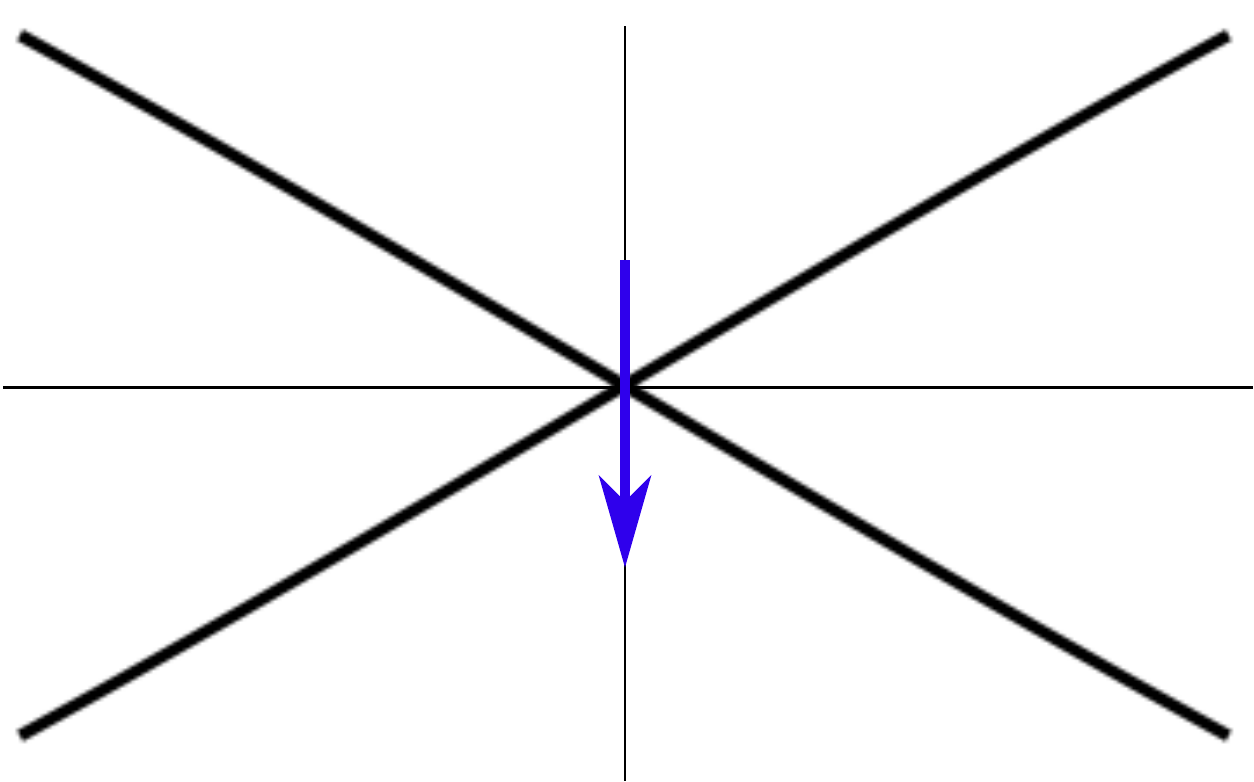}
		\caption{}
		\label{fig_quench_2}
	\end{subfigure}
	\begin{subfigure}{0.2\textwidth}
		\includegraphics[width=0.8\columnwidth]{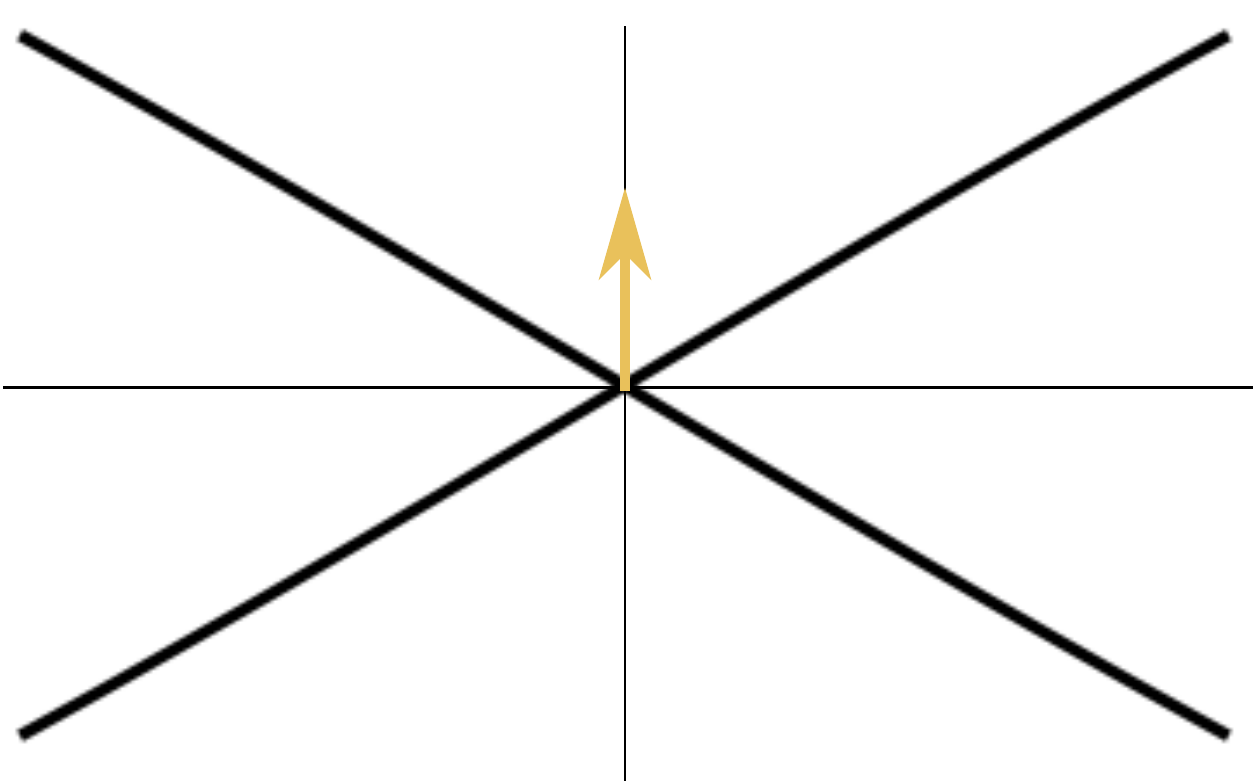}
		\caption{}
		\label{fig_quench_3}
	\end{subfigure}
	\begin{subfigure}{0.2\textwidth}
		\includegraphics[width=0.8\columnwidth]{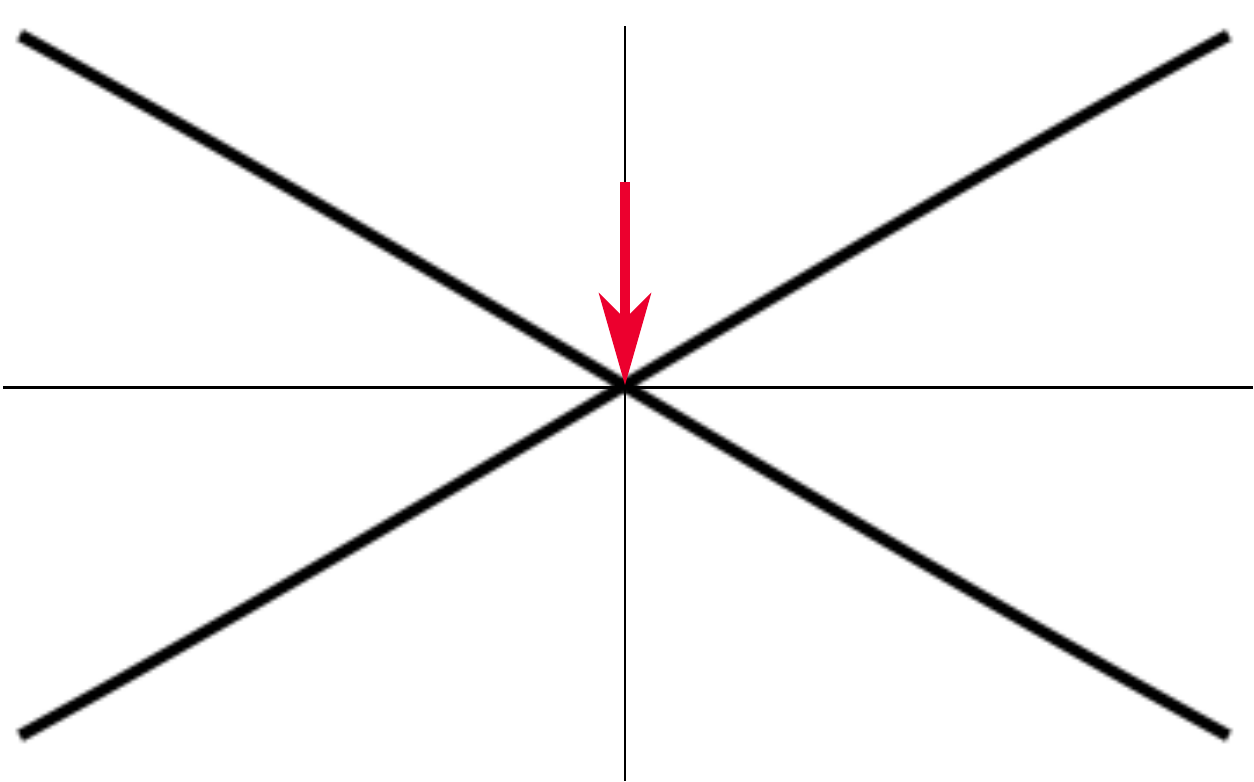}
		\caption{}
		\label{fig_quench_4}
	\end{subfigure}
	\cc
	\caption{Schematic of the quenches performed in the topologically trivial ($\phi=0$) massive graphene Hamiltonian near $M\approx 0$. The solid lines are the QCLs which are approximately linear for small $\phi$ and the arrows denote the direction of quench. Four cases are analyzed \textit{viz}., quench (a) without crossing QCP at ($M=0$), (b) across the  QCP, (c) from the QCP and (d) ending at the QCP. }
	\label{fig_triv}
\end{figure*}
\section{Casimir Free Energy}\label{sec_casimir}
As discussed already,  the universality in the behavior of $P(W)$ is directly linked to large $S$ behavior of $f_c(S)$. To present the outline of the procedure to extract  $f_c(S)$ from the total free energy, 
 we first substitute Eq.~\eqref{eq_gs} in Eq.~\eqref{eq_free_energy}  so that 
\begin{equation}
\log{Z(S)}=-L^2\left(2f_s + f_c(S)\right).
\label{eq_zs_log}
\end{equation}
where we have set $d=2$ for the $2$D Haldane model. The fact that the quasi-momentum modes are conserved and independent of each other, allows one to construct the initial state as
\begin{equation}
\ket{\psi_0}=\bigotimes_{\vec{k}}\ket{\psi_0({\vec{k}})}
\end{equation}
where $\ket{\psi_0(\vec{k})}$ is the  energy eigenstate of $H_i(\vk)$ and the direct product is taken over the first Brillouin zone (BZ) of the lattice. This simplification, together with Eq.~\eqref{eq_hk} immediately implies that Eq.~\eqref{eq_zs} can be rewritten as
\begin{equation}
Z(S)=e^{SE_f^0}\prod_{\vec{k}}\bra{\psi_0(\vec{k})}e^{-H_f(\vec{k})S}\ket{\psi_0(\vec{k})}.
\end{equation}
where $E_f^0=-\sum_{\vec{k}}\epsilon_f(\vec{k})$ and $-\epsilon_f(\vec{k})$ is the ground state energy of final Hamiltonian $H_f(\vec{k})$.

Further, Eq.~\eqref{eq_hk} also suggests that the Hilbert space of the decoupled two-level systems can be mapped to the surface of a Bloch sphere of radius $|\vec{h}|$. Let us assume that the initial state lies at a point ($\theta_i,\Phi_i$) on this Bloch sphere where $\theta$ and $\Phi$ are the azimuthal and polar angles, respectively. It can be easily checked from Eq.~\eqref{eq_hx}, \eqref{eq_hy} and \eqref{eq_hz} that the quench which is performed on $M$ only effects the $h_z(\vec{k})=|\vec{h}|\cos{\theta}$ component of $\vec{h}(\vec{k})$, thereby limiting the subsequent dynamics of the state to a great circle passing through the poles on the surface of the Bloch sphere. Finally, expanding $\psi_0(\vec{k})$ in the eigenbasis of $H_f(\vec{k})$, we obtain
\begin{equation}
Z(S)=\prod_{\vec{k}} \cos^2{(\varphi(\vec{k}))}\left(1+\tan^2{(\varphi(\vec{k}))}e^{-2S\epsilon_f(\vec{k})}\right)
\end{equation}  
where $\varphi(\vec{k})=\frac{\theta_f(\vec{k})-\theta_i(\vec{k})}{2}$ and $\theta_{i(f)}=\cos^{-1}{\frac{h_{z,i(f)}}{|\vec{h}_{i,(f)}|}}$. 

Substituting this expression for $Z(S)$ in Eq.~\eqref{eq_zs_log}, we have
\begin{multline}
-L^2\left(2f_s + f_c(S)\right)=\sum_{\vk}2\log{\left(\cos{(\varphi(\vk))}\right)}\\+\sum_{\vk}\log{\left(1+\tan^2{(\varphi(\vec{k}))}e^{-2S\epsilon_f(\vec{k})}\right)}.
\end{multline}
Now, assuming the continuum limit,  we can identify the surface and Casimir free energy contributions as
\begin{subequations}
	\begin{equation}\label{eq_fs}
	f_s=-\frac{1}{L^2A_{B}}\int_{BZ}\log{\left(\cos{(\varphi(\vk))}\right)}d\vk
	\end{equation}
	\begin{equation}\label{eq_fc}
	f_c(S)=-\frac{1}{L^2A_{B}}\int_{BZ}\log{\left(1+\tan^2{(\varphi(\vec{k}))}e^{-2S\epsilon_f(\vec{k})}\right)}d\vk
	\end{equation}
\end{subequations}
where $A_B=\int_{BZ}{\vec{dk}}$ is the area of the Brillouin zone. 
\section{Work statistics in topologically trivial graphene}\label{sec_triv}
When $\phi=0$, the amplitude of NNN hoppings are real and their only effect is to rescale the energy spectrum of the massive graphene Hamiltonian with NN hoppings by $h_0(\vk)$. We analyze the large $S$ behavior of $f_c(S)$ for quenches close to the gap-less graphene point ($M=0,\phi=0$ in Fig.~\ref{fig_haldane}) as follows:

It is clear from Eq.~\eqref{eq_fc} that in the large $S$ limit, the contributions to $f_c(S)$ from the quasi-momentum modes $\vk$ fall off exponentially as we move away from the two Dirac points $\vec{K}_1$ and $\vec{K}_2$  which are time-reversed partners of each other, thus  $\epsilon(\vec{K}_1)=\epsilon(\vec{K}_2)\approx0$ as $M$ tends to zero. Thus, the dominant contribution to the integral in Eq.~\eqref{eq_fc} comes from the lowest energy continuum around each of the Dirac points which contribute equally and identically to $f_c(S)$. Expanding the energy spectrum around $\vec{K}_1$ to leading non-trivial order in $k=|\vk-\vec{K}_1|$, we have
\begin{equation}
\epsilon(k)=\sqrt{M^2+k^2}.
\end{equation}   
In the continuum limit, the limits of the integration in Eq.~\eqref{eq_fc} extend to infinity to yield,
\begin{multline}\label{eq_fc_smallk}
f_c(S)=-\frac{1}{L^2A_{B}}\int_{-\infty}^{\infty}\int_{-\infty}^{\infty}dk_xdk_y\\\times\log{\left(1+\tan^2{(\varphi(\vec{k}))}e^{-2S\sqrt{M_f^2+k^2}}\right)}.
\end{multline} 
Further simplification requires the explicit form of $\tan{(\varphi(\vk))}$ for small $k$ which we now evaluate for several cases as elaborated below:

\subsection{Quench without crossing QCP ($M_i,M_f\gtrless0$)}\label{ssec_sp1}
In this case, the initial and the final Semenoff masses are either both positive or negative (Fig.~\ref{fig_quench_1}) and we have (see Appendix \ref{app_1})

\begin{equation}
\tan(\varphi(\vk))=C(M_i,M_f)k
\end{equation} 
to the leading order in $k$ where $C(M_i,M_f)=(M_i-M_f)/{2M_iM_f}$  depends only on $M_i$ and $M_f$. Substituting in Eq.~\eqref{eq_fc_smallk}, we get
\begin{multline}\label{eq_fc_1}
f^{1}_c(S)=-\frac{2\pi}{L^2A_{B}}\int_{0}^{\infty}\log\bigg(1+C^2(M_i,M_f) \\
	\times k^2e^{-2S\sqrt{M_f^2+k^2}}\bigg)kdk.
\end{multline} 
where the superscript $1$ in $f^{\vec{K}}_c(S)$ refers to the fact that we are  considering contribution from the lowest energy continuum from only around $\vec{K}_1$. Following few steps of algebra (see Appendix \ref{app_2} for detail), we eventually obtain
\begin{equation}\label{eq_fc_sp1}
f_c(S)=-2\times\frac{\pi (1-M_f/M_i)^2}{4L^2 A_B}\left(\frac{e^{-2S|M_f|}}{S^2}\right)
\end{equation}
where the multiplicating factor $2$ accounts for the fact that each Dirac point contributes identically. The characteristic function defined in Eq.~\eqref{eq_gs} takes the form
\begin{multline}
G(S)=e^{-\Delta E_0S}e^{-2L^2f_s}e^{-L^2 f_c(S)}\\=e^{-\Delta E_0S}e^{-2L^2f_s}\left(1-L^2f_c(S)+...\right)
\end{multline}
where we have expanded the third exponential to leading order in $f_c(S)$ exploiting the fact that $f_c(S)$ decays exponentially with $S$. Substituting the form of $f_c(S)$ from Eq.~\eqref{eq_fc_sp1} and performing an inverse Laplace transform on $G(S)$ finally gives us the small $W$ behavior of $P(W)$ as

\begin{multline}\label{eq_pw1}
P(W) =e^{-2L^2f_s}\bigg[\delta(W-\Delta E_0)+\Theta\left(W-\Delta E_0-2|M_f|\right)\\
\times\bigg\{\frac{\pi (1-M_f/M_i)^2}{2A_B}\left(W-\Delta E_0-2|M_f|\right)\bigg\}\bigg].
\end{multline}
$P(W)$ therefore has a delta function peak at $W=\Delta E_0$ and the presence of the Heavyside theta function in the second term implies the existence of an edge singularity. Note that the quench amplitudes and other microscopic details only appear in the coefficient of the edge-singularity while the exponent of $(W-\Delta E_0-2|M_f|)$ is independent of such details and is thus universal.  

\subsection{Quench across the QCP ($M_i\gtrless0\gtrless M_f$)}
When $M_i$ and $M_f$ are on either side of the gapless graphene point (Fig.~\ref{fig_quench_2}), the leading order term in the expansion of $\tan(\varphi(\vk))$ takes the form (again, referring to Appendix \ref{app_1})

\begin{equation}
\tan(\varphi(\vk))=-\frac{1}{C(M_i,M_f)k}.
\end{equation}
Proceeding similarly as in Case.~\ref{ssec_sp1}, we obtain the Casimir interaction term as
\begin{equation}\label{eq_fc_ap1}
f_c(S)=-\frac{16\pi(1-\gamma)M_i^2 M_f^2}{L^2 A_B(M_i-M_f)^2}e^{-2S|M_f|}
\end{equation}
 where $\gamma$ is the Euler-Mascheroni constant. The work distribution function is thus
\begin{multline}\label{eq_pw2}
P(W)= e^{-2L^2f_s}\bigg[\delta(W-\Delta E_0)+\delta\left(W-\Delta E_0-2|M_f|\right)\\
\times\frac{16\pi(1-\gamma)M_i^2 M_f^2}{ A_B(M_i-M_f)^2}\bigg]
\end{multline}
which interestingly has two delta function peaks at $W=\Delta E_0$ and $W=\Delta E_0+2|M_f|$ and contains no continuum.

\subsection{Quench from the QCP ($M_i=0$)}
If the quench originates from the critical (graphene) point (Fig.~\ref{fig_quench_3}), $\tan{(\varphi(\vk))}$ depends only on the relative position of $M_f$ and is independent of its absolute value.

\begin{equation}
\tan{(\varphi(\vk))}=-sgn(M_f)
\end{equation} 
The Casimir interaction term assumes the simple form
\begin{equation}
f_c(S)=-\frac{2\pi M_f}{L^2A_B}\left(\frac{e^{-2S|M_f|}}{S}\right)
\end{equation}
and the work distribution is 
\begin{multline}\label{eq_pw3}
P(W) =e^{-2L^2f_s}\bigg[\delta(W-\Delta E_0)\\
		+\frac{2\pi M_f}{A_B}\Theta\left(W-\Delta E_0-2|M_f|\right)\bigg]
\end{multline}
Thus, the continuum begins with a  finite discontinuity at $W=\Delta E_0+2|M_f|$ .

\subsection{Quench ending at the QCP($M_f=0$)}
In this case (Fig.~\ref{fig_quench_4}), $\tan{(\varphi(\vk))}$ once again is independent of the absolute value of $M_i$ and depends only on its relative position to the QCP.

\begin{equation}
\tan{(\varphi(\vk))}=sgn(M_i)
\end{equation}
However, $f_c(S)$ now undergoes a power law decay with $S$,
\begin{equation}
f_c(S)=-\frac{\pi}{L^2A_B}\left(\frac{1}{S^2}\right).
\end{equation}
This is expected from the fact that the correlation length diverges at the gapless critical point and the two-point correlations exhibit a power law decay. $P(W)$ thus assumes the form
\begin{multline}\label{eq_quench_4}
P(W)=e^{-2L^2f_s}\bigg[\delta(W-\Delta E_0)+\Theta\left(W-\Delta E_0\right)\\
	\times\bigg\{\frac{\pi}{L^2A_B}\left(W-\Delta E_0\right)\bigg\}\bigg],
\end{multline}
which shows that there is no gap in the low energy regime of $P(W)$ and the continuum starts  from $W=\Delta E_0$.
\begin{figure*}
	\begin{subfigure}{0.2\textwidth}
		\includegraphics[width=0.8\columnwidth]{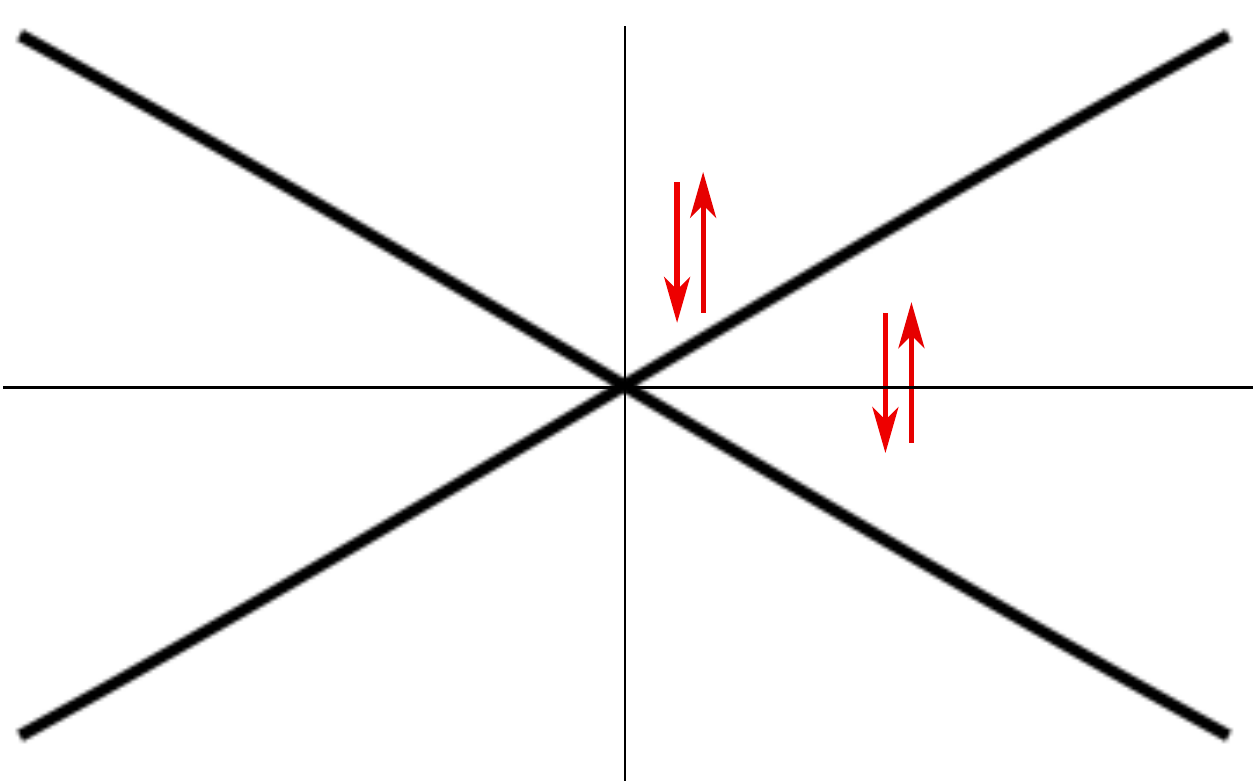}
		\caption{}
		\label{fig_quench_A}
	\end{subfigure}
	\begin{subfigure}{0.2\textwidth}
		\includegraphics[width=0.8\columnwidth]{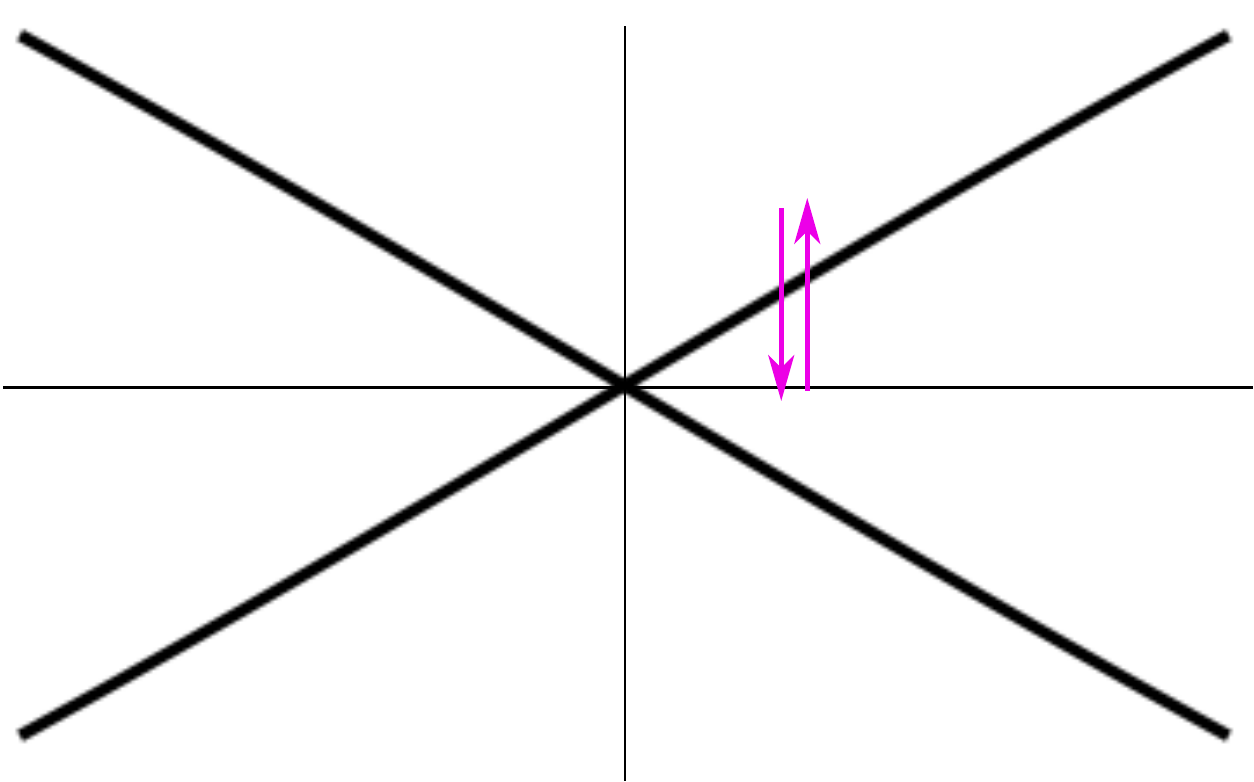}
		\caption{}
		\label{fig_quench_B}
	\end{subfigure}
	\begin{subfigure}{0.2\textwidth}
		\includegraphics[width=0.8\columnwidth]{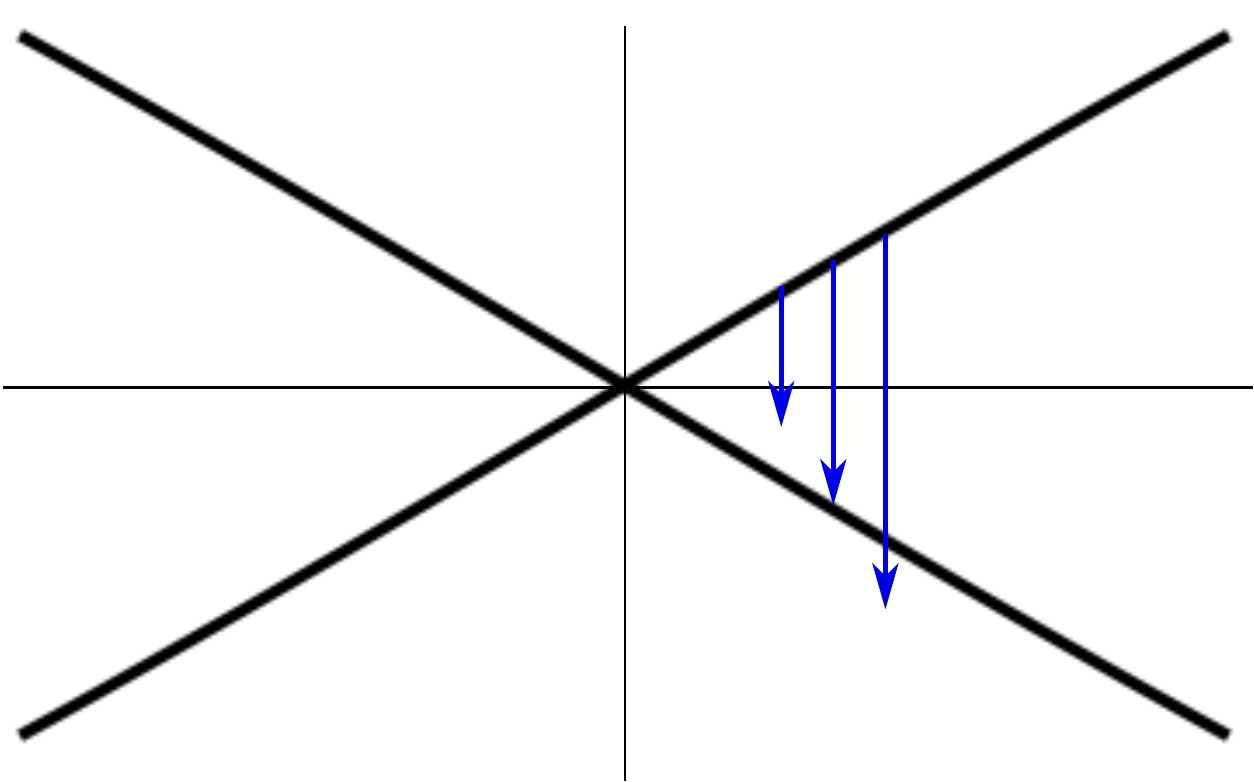}
		\caption{}
		\label{fig_quench_C}
	\end{subfigure}
	\begin{subfigure}{0.2\textwidth}
		\includegraphics[width=0.8\columnwidth]{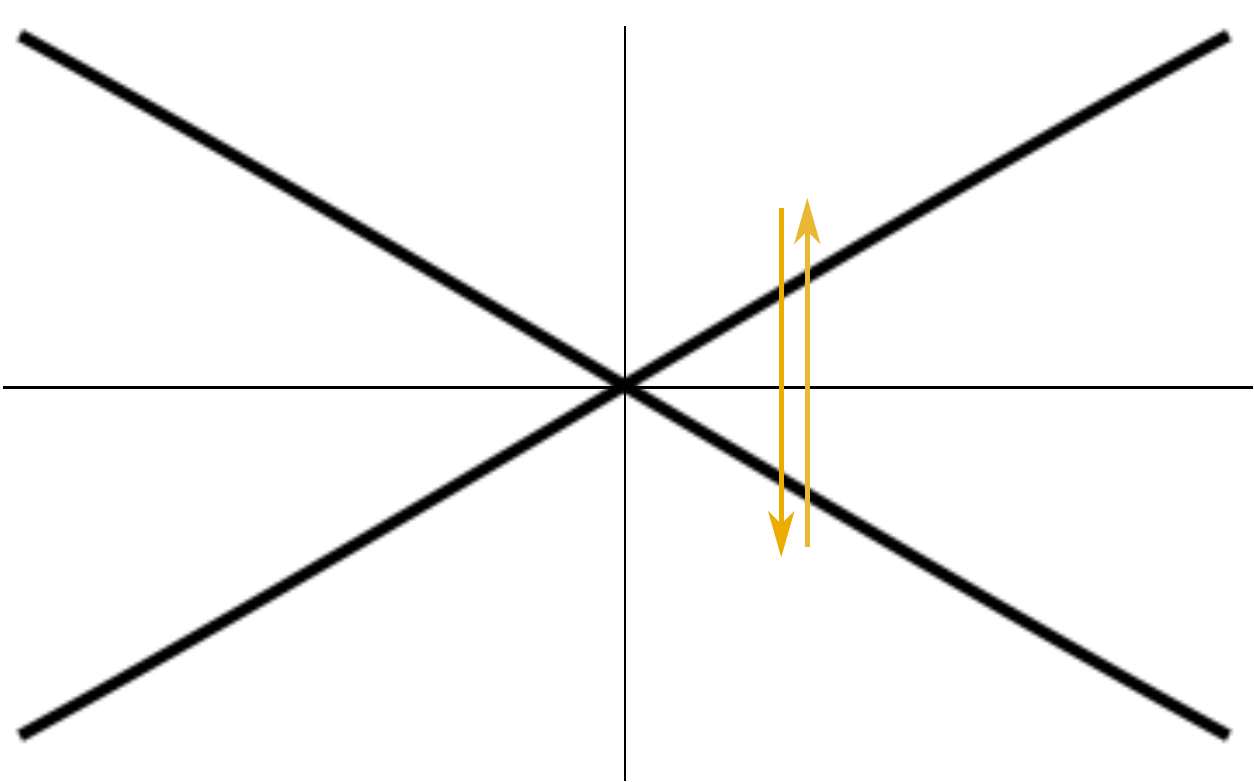}
		\caption{}
		\label{fig_quench_D}
	\end{subfigure}
	\cc
	\caption{Schematic of the quenches performed in the topological Haldane model with a small ($\phi\neq 0$) and near the QCLs. The cases analyzed are  quench (a) within same phase, (b) across one QCL, (c) starting from the QCL and (d) across the topological phase. }
\end{figure*}

In summary, we demonstrated the appearance of universality in the small $W$ limit of the work distribution function following a sudden quench in the topologically trivial graphene Hamiltonian. Intuitively, one can provide a physical interpretation for the terms appearing in $P(W)$ as follows. For quenches performed without crossing the QCP, the delta function term at $W=\Delta E_0$ in the Eq.~$\eqref{eq_pw1}$ corresponds to the reversible work done in the adiabatic limit. The Heavyside theta function term on the other hand indicates that the threshold for quasi particle excitations is equal to the minimum energy gap $2|M_f|$ in the spectrum of $H_f$. These excitations correspond to the irreversible work performed during the post quench dynamics. Since a quench performed across the QCP involves closing of the gap in the spectrum, excitations are possible even in the adiabatic limit  which explains the appearance of a second delta function term at $W=\Delta E_0 + 2|M_f|$ (see Eq.~$\eqref{eq_pw2}$). We would also like to point out that the finite discontinuity at the edge for quenches originating from the QCP (Eq.~\eqref{eq_pw3})did not appear in the case of the one dimensional Ising model and is associated with the higher dimensionality of our system which is two dimensional. Finally, the absence of any edge for quenches ending at the QCP (Eq.~\eqref{eq_quench_4}) is simply because of the fact that spectrum of the final Hamiltonian is gapless and no threshold exists for quasi particles excitations.

\section{Work statistics in topological Haldane model}\label{sec_topo}
In this section, we shall set $\phi\neq 0$ and probe the non-trivial influence of the equilibrium topology on the distribution function $P(W)$. Let us recall that  when complex NNN hoppings are introduced into the graphene Hamiltonian, the two Dirac points are no longer connected through TRS. The resulting asymmetry in the spectrum prohibits simultaneous gap-closings at the two Dirac points. The system now has two quantum critical lines (QCLs) (Fig.~\ref{fig_haldane}), $M_{c1}=3\sqrt{3}t'\sin{\phi}$ and $M_{c2}=-3\sqrt{3}t'\sin{\phi}$ for vanishing of the two Dirac points, respectively. However if $\phi$ is small, the two QCLs are very close to each other as
\begin{equation}\label{eq_QCP_diff}
 |M_{c1}-M_{c2}|\approx6\sqrt{3}t'|\phi|.
\end{equation}
Importantly, the spectrum at the two Dirac points, though non-identical, are still of the same orders of magnitude. For our purpose, this means that the lowest energy continuum around both the Dirac points still make dominant contributions to the Casimir interaction term $f_c(S)$ in the large $S$ limit. The spectrum around the Dirac points can now be expanded to leading non-trivial order in $k_1=|\vk-\vec{K}_1|$  and $k_2=|\vk-\vec{K}_2|$ in the form
	\begin{equation}
	\epsilon(k_{1(2)})=\sqrt{m_{1(2)}^2 + k_{1(2)}^2}
	\end{equation}
where $m_{1}=M-3\sqrt{3}t'\phi$ and $m_{2}=M+3\sqrt{3}t'\phi$. A quench in the Semenoff mass from $M_i$ to $M_f$ is therefore equivalent to simultaneous quenches in $m_{1}$ and $m_{2}$ from ($m_{1i}, m_{2i}$) to ($m_{1f},m_{2f}$). In view of the above situations, we now proceed to evaluate $f_c(S)$ and $P(W)$ for the following cases:  

\subsection{Quench within trivial phase ($M_i,M_f\gtrless\pm3\sqrt{3}t'\sin{\phi}$) or within topological phase ($3\sqrt{3}t'\sin{\phi}>M_i,M_f>-3\sqrt{3}t'\sin{\phi}$)}
The situation here (Fig.~\ref{fig_quench_A}) is similar to the quenches carried out without crossing QCP in the massive graphene model. We observe that
\begin{subequations}
\begin{equation}
\tan(\varphi(k_1))=C(m_{1i},m_{1f})k_1
\end{equation}
\begin{equation}
\tan(\varphi(k_2))=C(m_{2i},m_{2f})k_2.
\end{equation}
\end{subequations}
The  Casimir interaction term assumes the form
\begin{multline}
f_c(S)=-\frac{\pi }{4L^2 A_B}\bigg[(1-m_{1f}/m_{1i})^2\frac{e^{-2S|m_{1f}|}}{S^2}\\+(1-m_{2f}/m_{2i})^2\frac{e^{-2S|m_{2f}|}}{S^2}\bigg].
\end{multline} 
and the work distribution function is obtained as
\begin{multline}
P(W) =e^{-2L^2f_s}\bigg[\delta(W-\Delta E_0)+\Theta\left(W-\Delta E_0-2|m_{1f}|\right)\\
\times\bigg\{\frac{\pi (1-m_{1f}/m_{1i})^2}{4A_B}\left(W-\Delta E_0-2|m_{1f}|\right)\bigg\}\\
+\Theta\left(W-\Delta E_0-2|m_{2f}|\right)\\
\times\bigg\{\frac{\pi (1-m_{2f}/m_{2i})^2}{4A_B}\left(W-\Delta E_0-2|m_{2f}|\right)\bigg\}\bigg].
\end{multline}

Comparing with Eq.~\eqref{eq_pw1}, we see that $P(W)$ now has two Heavyside theta funIctions indicating the existence of two different thresholds for quasi particle excitations. This is a consequence of the unequal energy gaps at the two Dirac points resulting from broken TRS.
\newcolumntype{C}[1]{>{\centering\let\newline\\\arraybackslash\hspace{0pt}}m{#1}}
\begin{table*}
	
	\begin{tabular}{|C{3cm}|C{2cm}|C{2cm}|C{2cm}|C{4cm}|}
		\hline
		Quench & Additional delta-function peak position(s) at $W=$ & Theta function discontinuity position(s) at $W=$& Scaling exponent of $W$ associated with Theta function edge  &  Overall nature of $P(W)$ for small $W$\\
		\hline
		A. within trivial or within topological phase & - & i. $2|m_{1f}|$\newline ii. $2|m_{2f}|$ & i. $1$ \newline ii. $1$ & Continuum starts from $min\{2|m_{1f}|,2|m_{2f}|\}$ and the slope changes sharply at $max\{2|m_{1f}|,2|m_{2f}|\}$. \\
		\hline
		B. from trivial to topological phase or vice-versa & $2|m_{1f}|$ & $2|m_{2f}|$ & $1$ & Continuum starts from $2|m_{2f}|$ and a delta function peak exist at $2|m_{1f}|$, which may either lie prior to the continuum or be superimposed on it.\\
		\hline
		C. away from one QCL and ending:\newline\vspace{1cm} a. before the other QCL\newline\vspace{1cm} b. on the other QCL\newline\vspace{1cm} c. across the other QCL  & a. - \newline\vspace{2.4cm} b. -\newline\vspace{2.4cm} c. $2|m_{2f}|$& a.i. $2|m_{1f}|$\newline a.ii. $2|m_{2f}|$\newline\vspace{2 cm} b.i. $0$\newline b.ii. $2|m_{1f}|$\newline\vspace{2cm} c.  $2|m_{1f}|$  & a.i. 0\newline a.ii.$1$ \newline\vspace{2cm}b.i.$1$\newline b.ii. 0\newline\vspace{2cm}c. 0& a. If $|m_{1f}|\leq |m_{2f}|$, continuum starts with a non-zero finite value at $2|m_{1f}|$ and the slope changes sharply at $2|m_{2f}|$; if $|m_{1f}|> |m_{2f}|$, continuum begins at $2|m_{2f}|$ with a finite discontinuity at $2|m_{1f}|$ \newline\vspace{0.3cm} b. Continuum starts from the origin and the slope changes sharply at $2|m_{1f}|$.\newline\vspace{0.3cm} c. Continuum starts from $2|m_{1f}|$ with a non-zero finite value and a delta function peak exist at $2|m_{2f}|$, which may either lie prior to the continuum or be superimposed on it.\\ 
		\hline
		D. across the topological phase & i. $2|m_{1f}|$\newline ii. $2|m_{2f}|$ & - & - & Delta function peaks at $2|m_{1f}|$ and $2|m_{2f}|$.\\
		\hline
	\end{tabular}
	\cc
	\captionof{table}{Summary of the universal characteristics of $P(W)$ for quenches performed in the topological Haldane model. $W$ has been rescaled to $W=W-\Delta E_0$.  In all the cases, there is a delta function at $W=0$ which has not been reported separately here.}
	\label{tab_sum}
\end{table*}
\subsection{Quench from trivial to topological phase ($M_i\gtrless\pm3\sqrt{3}t'\sin{\phi}\gtrless M_f\gtrless\mp 3\sqrt{3}t'\sin{\phi}$) or vice-versa} \label{subsec_B}
In this case (Fig.~\ref{fig_quench_B}), the quench is performed across one of the two QCLs. One finds:
\begin{subequations}
\begin{equation}
\tan(\varphi(k_1))=-\frac{1}{C(m_{1i},m_{1f})k_1}
\end{equation}
\begin{equation}
\tan(\varphi(k_2))=C(m_{2i},m_{2f})k_2.
\end{equation}
\end{subequations}
It should be noted that unlike the previous case, $\tan(\varphi(k))$ has a pole at $k_1=0$ while it is analytic for $k_2$. Therefore, the two Dirac points contribute differently to the Casimir interaction term and we obtain:
 \begin{multline}
f_c(S)=-\frac{\pi}{L^2A_B}\bigg[\frac{8(1-\gamma)m_{1i}^2 m_{1f}^2}{(m_{1i}-m_{1f})^2}e^{-2S|m_{1f}|}\\
		+(1-m_{2f}/m_{2i})^2\frac{e^{-2S|m_{2f}|}}{4S^2}\bigg].
\end{multline} 
This is also reflected in the work distribution as
\begin{multline}
P(W)= e^{-2L^2f_s}\bigg[\delta(W-\Delta E_0)+\Theta\left(W-\Delta E_0-2|m_{2f}|\right)\\
\times\bigg\{\frac{\pi (1-m_{2f}/m_{2i})^2}{4A_B}\left(W-\Delta E_0-2|m_{2f}|\right)\bigg\}\\+\frac{8\pi(1-\gamma)m_{1i}^2 m_{1f}^2}{ A_B(m_{1i}-m_{1f})^2}\delta\left(W-\Delta E_0-2|m_{1f}|\right)\bigg].
\end{multline}
We notice that there now exist both a delta function term and a Heavyside theta function in the leading order. Particularly, if $|m_{2f}|<|m_{1f}|$, an adiabatic contribution will be superimposed on the quasi particle continuum in $P(W)$ after the edge. This is a non trivial behavior which does not occur for quenches in the trivial phase.

\subsection{Quench starting from the QCLs($M_i=\pm3\sqrt{3}t'\sin{\phi}$)}
For quenches originating from one of the QCLs, there are three possible scenarios (Fig.~\ref{fig_quench_C}) depending on relative position of $M_f$ with respect to the other QCL. For example, if the quench originates from $M_i=3\sqrt{3}t'\phi$, the Casimir interaction term and the work distribution function for each of the three scenarios are listed below:
\paragraph{$\underline{M_f>-3\sqrt{3}t'\phi}$}
\begin{subequations}
\begin{multline}
f_c(S)=-\frac{\pi}{L^2A_B}\bigg[m_{1f}\frac{e^{-2S|m_{1f}|}}{S}\\
		+\left(1-\frac{m_{2f}}{6\sqrt{3}t'\sin{\phi}}\right)^2\frac{e^{-2S|m_{2f}|}}{4S^2}\bigg]
\end{multline}
\begin{multline}
P(W)=e^{-2L^2f_s}\bigg[\delta(W-\Delta E_0)\\
+\frac{\pi m_{1f}}{A_B}\Theta\left(W-\Delta E_0-2|m_{1f}|\right)+\Theta\left(W-\Delta E_0-2|m_{2f}|\right)\\
\times\bigg\{\frac{\pi}{4A_B}\left(1-\frac{m_{2f}}{6\sqrt{3}t'\sin{\phi}}\right)^2\left(W-\Delta E_0-2|m_{2f}|\right)\bigg\}\bigg]
\end{multline}
\end{subequations}	
Here, $P(W)$ consists two Heavyside theta functions and there exists  a finite discontinuity at $W=2|m_{1f}|$.

\paragraph{$\underline{M_f=-3\sqrt{3}t'\sin{\phi}}$}
\begin{subequations}
\begin{equation}
f_c(S)=-\frac{\pi}{L^2A_B}\bigg[m_{1f}\frac{e^{-2S|m_{1f}|}}{S}+\frac{1}{2S^2}\bigg]
\end{equation}

\begin{multline}
P(W)=e^{-2L^2f_s}\bigg[\delta(W-\Delta E_0)
		+\Theta\left(W-\Delta E_0\right)\\
		\times\frac{\pi}{L^2A_B}\left(W-\Delta E_0\right)+\frac{\pi m_{1f}}{A_B}\Theta\left(W-\Delta E_0-2|m_{1f}|\right)\bigg]
\end{multline}
\end{subequations}
Once again, we obtain two Heavyside theta functions and continuum begins from $W=\Delta E_0$ with no gapped region.
\paragraph{$\underline{M_f<-3\sqrt{3}t'\sin{\phi}}$}
\begin{subequations}
\begin{multline}
 f_c(S)=-\frac{\pi}{L^2A_B}\bigg[m_{1f}\frac{e^{-2S|m_{1f}|}}{S}\\+\frac{8(1-\gamma)m_{2i}^2 m_{2f}^2}{(m_{2i}-m_{2f})^2}e^{-2S|m_{2f}|}\bigg]
\end{multline}
\begin{multline}
 P(W)=e^{-2L^2f_s}\bigg[\delta(W-\Delta E_0)\\
 		+\frac{\pi m_{1f}}{A_B}\Theta\left(W-\Delta E_0-2|m_{1f}|\right)+\delta\left(W-\Delta E_0-2|m_{2f}|\right)\\\times\frac{8\pi(1-\gamma)m_{2i}^2 m_{2f}^2}{ A_B(m_{2i}-m_{2f})^2}\bigg]
\end{multline}
\end{subequations}
Here,  $P(W)$ has a delta function peak at $W=2m_{2f}$ and the continuum begins with a finite discontinuity. It is evident that in all the above subcases, the resulting $P(W)$ shows multiple thresholds as well as additional adiabatic contributions similar to those obtained in case \ref{subsec_B}. 
 
\subsection{Quench across the topological phase ($M_i>3\sqrt{3}t'\sin{\phi},M_f<-3\sqrt{3}t'\sin{\phi}$)}
The quench in this case is performed across the topological phase from one trivial phase to other as indicated in Fig.~\ref{fig_quench_D}. Proceeding as before, the work distribution function evaluates to

\begin{multline}
P(W)= e^{-2L^2f_s}\bigg[\delta(W-\Delta E_0)+\\
+\frac{8\pi(1-\gamma)}{A_B)}\bigg\{\frac{m_{1i}^2 m_{1f}^2}{(m_{1i}-m_{1f})^2}\delta\left(W-\Delta E_0-2|m_{1f}|\right)\\
+\frac{m_{2i}^2 m_{2f}^2}{ (m_{2i}-m_{2f})^2}\delta\left(W-\Delta E_0-2|m_{2f}|\right)\bigg\}\bigg].	
\end{multline} 
Therefore, there exist two additional delta function peaks and no continuum. 


\section{Discussions and Conclusions}\label{sec_dis}
Let us now summarize our results as follows. We outlined the universalities in the work distribution function for the case of topologically trivial graphene Hamiltonian in Sec.~\ref{sec_triv} . TRS ensured that the contributions from the two Dirac points were identical and the expressions for$P(W)$
 thus obtained are similar to those available in literature for the one dimensional transverse Ising model \cite{gambassi_ar_11, smacchia13}. Next, on introducing a small non-zero value for $\phi$, the TRS is broken and the resulting inequivalent spectrum at the two Dirac points leads to emergence of new behavior like the existence of multiple thresholds for quasi particle excitations and  superimposition of adiabatic contributions on the continuum of irreversible excitations, as summarized in  Table.~\ref{tab_sum}. However, we would like to point out that for quenches performed in $M$ at a large constant value of $\phi$, the  approximate equality in Eq.~\eqref{eq_QCP_diff} is no longer satisfied. The spectrum at the two Dirac points are of different orders of magnitude and therefore only one of them contributes dominantly to the Casimir interaction term for any given quench. In this scenario, the $P(W)$ reduces to a form similar to that obtained in the topologically trivial case where the two Dirac points contributed identically.
 
We can therefore conclude that although the work distribution function in general displays similar universal behavior for the topologically trivial graphene and the topological Haldane model in the $W\to 0$ limit for large values of $\phi$, it may however acquire a new class of universal behavior for quenches performed arbitrarily close to $\phi=0$. We again note here that the breaking of the TRS which endows topological structure to the graphene Hamiltonian, is also at the root of the emergence of these new behaviors; hence signifying that the system's equilibrium topology may have a direct bearing on the work distribution function at least for some values of system parameters. This is significant because the work distribution function now exhibits different universal behaviors following the non-equilibrium dynamics of the system for small $\phi$ and large $\phi$ limits, although the two limits belong to the same equilibrium universality class as far as our system is concerned for all non-zero values of (TRS breaking) $\phi$.

\section{Experimental Possibilities}\label{sec_exp}
Extracting the work distribution function or its characteristic function is not an easy task experimentally as it  requires two projective and non destructive measurements on the eigenbases of initial and final Hamiltonians. However, significant progress has been made in recent times, noticeable of which are the use of Ramsey interferometry on an ancillary qubit for extraction of the characteristic function \ct{dorner13, mazzola13} and the extraction of the work distribution function  with Rubidium atoms on an atom chip \ct{cerisola17}. The former technique has been used to verify fluctuation theorems for a quantum system in a nuclear magnetic resonance platform \cite{batalhao14}. The Haldane model on the other hand, which is the system that we have considered in our work, has also been realized experimentally by preparing  non-interacting ultracold fermionic gas on an optical honeycomb lattice \ct{jotzu14}. TRS is broken though circular modulation of the lattice positions while a magnetic field gradient effectively plays the role of the Semenoff mass. Since all the quenches considered in our work are on the Semenoff mass only, experimentally verifying our results with the ancillary qubit technique using Ramsey interferometry will only require a quench of the magnetic field gradient with constant periodic modulation of the lattice position. We therefore believe that the experimental verification of our results, although difficult to achieve, will be possible.

\section*{Acknowledgement}
AD acknowledges SERB, DST, New Delhi for financial support. SB acknowledges CSIR, India for financial sup- port. We also acknowledge Souvik Bandyopadhyay, Sudarshana Laha and Somnath Maity for their critical comments.
\appendix

\section{Evaluation of $\tan{(\varphi(\vk))}$ for small $\vk$ for trivially gapped graphene}\label{app_1}
We have,
\begin{subequations}\label{eq_trig}
\begin{equation}
\cos{\theta(\vk)}=\frac{h_z(\vk)}{\ek}=M/\epsilon(k)
\end{equation}
where $h_z(\vk)$ is actually independent of $\vk$ for each Dirac point in all the cases we consider throughout.  
\begin{equation}
\sin{\theta(\vk)}=\frac{\sqrt{h_x^2(\vk)+h_y^2(\vk)}}{\ek}\approx\frac{k}{\ek}
\end{equation}
\end{subequations}
to leading order in $k$.
A simple trigonometric manipulation allows one to write
\begin{equation}
\tan{(\varphi(\vk))}=\frac{1-\cos{(\theta_f(\vk)-\theta_i(\vk))}}{\sin{(\theta_f(\vk)-\theta_i(\vk))}}.
\end{equation}
Substituting Eq.~\eqref{eq_trig} in the above equation, we obtain
\begin{equation}\label{eqapp_tan}
\tan{(\varphi(\vk))}=\frac{\sqrt{(M_f^2+k^2)(M_i^2+k^2)}-M_iM_f-k^2}{(M_i-M_f)k}
\end{equation}
Expanding binomially and retaining terms upto $O[k^2]$,
\begin{equation}
\tan{(\varphi(\vk))}=\frac{|M_i||M_f|-M_iM_f+k^2\frac{(M_i-M_f)^2}{2M_iM_f}}{(M_i-M_f)k}
\end{equation}
Hence if $M_i,M_f\gtrless0$, we have
\begin{equation}
\tan{(\varphi(\vk))}=\frac{(M_i-M_f)}{2M_iM_f}k
\end{equation} 
where we have retained only the leading order term in $k$. Similarly, if $M_i\gtrless 0\gtrless M_f$, the leading order term is
\begin{equation}
\tan{(\varphi(\vk))}=-\frac{2M_iM_f}{(M_i-M_f)k}.
\end{equation} 
Finally, it is straightforward to see from Eq.~\eqref{eqapp_tan} that if  $M_i=0$\big($M_f=0$\big), we have $\tan{(\varphi(\vk))}=-sgn(M_f)$\big($sgn(M_i)$\big) respectively.

\section{Evaluation of the integral form of the Casimir term}\label{app_2}
We choose $\tan{(\varphi(\vk))}=C(M_i,M_f)k$ to outline the procedure for evaluating $f_c(S)$. Other forms of $\tan{(\varphi(\vk))}$ can be likewise evaluated. First we recall the following inverse Mellin transformation,
\begin{equation}
\log{(1+x)}=\frac{1}{2\pi i}\int_{a-i\infty}^{a+i\infty}\frac{\pi}{u\sin{\pi u}}x^{-u}du
\end{equation}
where $u\in\mathbb{C}$ and $-1<a<0$. Substituting in Eq.~\eqref{eq_fc_smallk},
\begin{multline}\label{eqapp_fc2}
f_c^1(S)=\frac{i}{L^2A_B}\int_{0}^{\infty}kdk\\
\times\int_{a-i\infty}^{a+i\infty}\frac{\pi}{u\sin{\pi u}}C(M_i,M_f)^{-2u}k^{-2u}e^{2uS|M_f|}e^{uSk^2/|M_f|}du
\end{multline}
where we have expanded $\epsilon_f(k)$ to order $O[k^2]$ as
\begin{equation}
\epsilon_f(k)=\sqrt{M_f^2+k^2}=|M_f|(1+k^2/2M_f^2)
\end{equation} 
The integral in $k$ can be evaluated as $Re[u]=a<0$, and therefore the Eq.~\eqref{eqapp_fc2} assumes the form
\begin{multline}
f_c^1(S)=\frac{i}{L^2A_B}\int_{a-i\infty}^{a+i\infty}\frac{\pi(-u)^u\Gamma(-u)}{2u\sin{\pi u}}\left(\frac{S}{M_f}\right)^{u-1}\\C(M_i,M_f)^{-2u}e^{2uS|M_f|}du\\
        =\frac{i}{L^2A_B}\int_{a-i\infty}^{a+i\infty}g(u)du
\end{multline}
Further,
\begin{equation}\label{eqapp_gu}
\int_{a-i\infty}^{a+i\infty}g(u)du=\int_{b-i\infty}^{b+i\infty}g(u)du+\sum_{b<Re[u]<a}res[g(u)]
\end{equation}
where $b<a$ and the residues are summed up over all the poles that lie within the strip $b<Re[u]<a$. The integrand $g(u)$ has poles on the real axis, which can be easily seen if we notice that
\begin{equation}
\frac{\pi}{\sin{\pi u}}=\Gamma(u)\Gamma(1-u),
\end{equation}
and the gamma function has  simple poles at $u=-n$ where $n\in\mathbb{I}^+$. The residue at the $n^{th}$ pole is  $(-1)^n/n!$. On choosing $b=-\infty$, the integral in the R.H.S. of Eq.~\eqref{eqapp_gu} reduces to zero and the summation is now over all $n\in\mathbb{I}^+$. However, since $S$ is large, we consider only the contribution from the pole at $u=-1$ and therefore we obtain,
\begin{multline}
f_c^1(S)=\frac{i}{L^2A_B}\frac{\Gamma(1)\Gamma(2)}{-2}\left(\frac{S}{M_f}\right)^{-2}C(M_i,M_f)^{2}e^{-2S|M_f|}\\\times(-2\pi i)
\end{multline}
where the last term within braces is the residue of $\Gamma(-1)$, \textit{i.e.}  $-1$, multiplied by $2\pi i$. The final expression is therefore,
\begin{equation}
f_c^1(S)=-\frac{\pi (1-M_f/M_i)^2}{4L^2 A_B}\left(\frac{e^{-2S|M_f|}}{S^2}\right).
\end{equation}

\end{document}